\documentclass[preprint,showpacs,preprintnumbers,amsmath,amssymb]{revtex4}

\usepackage{graphicx}% Include figure files
\usepackage{dcolumn}% Align table columns on decimal point
\usepackage{bm}% bold math

\begin{document}

\preprint{ANL-HEP-PR-02-020}
\preprint{FERMILAB-PUB-02/032-T}
\preprint{hep-ph/0202176}

\title{Next-to-leading order corrections to\\
$W+2$~jet and $Z+2$~jet production at hadron colliders} 

\author{John Campbell}
\email{johnmc@hep.anl.gov}
\affiliation{
HEP Division, Argonne National Laboratory,\\
9700 South Cass Avenue, Argonne, IL 60439}

\author{R. K. Ellis}%
\email{ellis@fnal.gov}
\affiliation{
Theory Group, Fermi National Accelerator Laboratory,\\
P.O. Box 500, Batavia, IL 60510}

\date{February 18, 2002}

\begin{abstract}
We report on QCD radiative corrections to the processes 
$p {\bar p} \to W + \mbox{2~jets}$ and
$p {\bar p} \to Z +\mbox{2~jets}$ at the Tevatron. These processes
are included in the Monte Carlo program MCFM, which allows the 
calculation of any infra-red finite variable at next-to-leading
order. Due to a better
theoretical description of jets at next-to-leading order, 
some distributions exhibit significant corrections.
As expected, the unphysical dependence of theoretical
predictions upon the renormalization and factorization scales
is greatly reduced compared to leading order. 
As an example of the predictions that may now be made with MCFM,
we present a next-to-leading order estimate of the heavy flavor
content of jets produced in association with vector bosons.
\end{abstract}

% PACS, the Physics and Astronomy
                             % Classification Scheme.
%\keywords{Suggested keywords}%Use showkeys class option if keyword
                              %display desired
\maketitle

\def\beq{\begin{equation}}
\def\eeq{\end{equation}}
\def\beqn{\begin{eqnarray}}
\def\eeqn{\end{eqnarray}}
\def\ran{\rangle}
\def\lan{\langle}
\def\bea{\begin{eqnarray}}
\def\eea{\end{eqnarray}}
\def\rar{\rightarrow}
\def\lar{\longrightarrow}
\def\lra{\leftrightarrow}
\def\ovl{\tilde}
\def\ie{{\it i.e.}}
\def\GeV{{\;\rm GeV}}
\def\MeV{{\;\rm MeV}}
\def\TeV{{\;\rm TeV}}
\def\blim{b_{\;\rm lim}}
\def\as{\alpha_S}
\def\bas{\bar{\alpha}_S}

\section{Introduction}

In this paper we report on the results of a calculation of the 
next-to-leading order QCD corrections to the processes,
\begin{eqnarray}
p+\bar{p} & \rightarrow & W + \mbox{2~jets},
\nonumber \\
p+\bar{p} & \rightarrow & Z/\gamma^* + \mbox{2~jets}.
\end{eqnarray}
These reactions will be investigated 
at the Tevatron, \ie~$p \bar{p}$ collisions at $\sqrt{s}=2$~TeV.
Our calculations are equally applicable to 
the LHC ($pp$ collisions at $\sqrt{s}=14$~TeV).
We plan to consider the LHC in a subsequent publication.
The results are obtained from the Monte Carlo program MCFM which allows
us to obtain full predictions for any infra-red safe variable.
In order to obtain fully differential distributions,
to which experimental cuts may be applied, various decay
modes of the $Z/\gamma^*$ intermediate states are included,
\begin{eqnarray}
      Z/\gamma^* & \to & e^- e^+ \nonumber \\
      Z/\gamma^* & \to & b \bar{b} \nonumber \\
      Z & \to & \sum_{i} \nu_i \bar{\nu}_i 
\end{eqnarray}
as well as,
\begin{eqnarray}
      W^+ \to & \nu_e e^+ 
\end{eqnarray}
for the $W^+$ decay. In this paper we will only report on leptonic decays of
the vector bosons. We use the approximation of massless leptons, so that our
results are also valid for the decays
$W^+ \to \nu_\mu \mu^+, W^+ \to \nu_\tau \tau^+$ in this approach.

Because of their phenomenological importance,
processes involving 
the production of vector bosons and jets have been considered by many
authors. $W$-boson production with two jets was considered at leading order 
in ref.~\cite{Kleiss:1985yh,Ellis:1985vn,Mangano:1989gs}. 
The same process involving jets at large rapidity
was considered in ref.~\cite{Andersen:2001ja}.
Vector boson production in association with $n$-jets for $n \leq 4$
is calculated at leading order in refs.~\cite{Berends:1990ax,Berends:1989cf}.
In refs.~\cite{Ellis:1981hk,Arnold:1989dp,Arnold:1989ub,Giele:dj}
predictions were made for processes involving a vector boson recoiling
against one jet at next-to-leading order.
In refs.~\cite{Giele:1996aa,Giele:1995kr} predictions were made for processes
involving $W$ bosons and one heavy quark at next-to-leading order. 
However to the best of our knowledge this is the first paper to calculate
vector boson processes with two jets at next-to-leading order.

In performing these calculations we have used the results
of other authors for the crossed reactions
$e^+ e^- \to \mbox{4 partons}$~\cite{Bern:1997sc}  
and $e^+ e^- \to \mbox{5 partons}$~\cite{Nagy:1998bb}.  
Even with the amplitudes in hand, the 
implementation in a Monte Carlo program requires considerable effort.

In order to highlight the similarities between the effects of
radiative corrections on the $W/Z + 2~{\rm jet}$ rate and 
the $W/Z + 1~{\rm jet}$ rate, we will also present some results for
the latter process. Such corrections have been known for some
time~\cite{Giele:dj} and have provided an invaluable tool for studies
at the Tevatron. The inclusion of the $W/Z + 1~{\rm jet}$ processes in
MCFM was useful to understand the issues to be faced in implementing
the more complicated $W/Z + 2~{\rm jet}$ processes.

We will also tie together our results with previous predictions made using
MCFM~\cite{Ellis:1998fv,Campbell:2000bg} to provide a 
consistent next-to-leading order prediction for the heavy flavor 
content of jets produced in association with a $W/Z$. Specifically
we consider vector boson events containing two tagged $b$-jets.
This quantity will be used in assessing the backgrounds to a number 
of new physics searches at the Tevatron~\cite{Carena:2000yx}. 
Experimental studies~\cite{Acosta:2001ct} have so 
far relied upon leading order predictions
as theoretical input.

\section{$\mbox{Hadron} + \mbox{Hadron} \rightarrow W/Z + \mbox{2~jets}$}

\begin{figure}
\centering
\includegraphics[width=8.6cm]{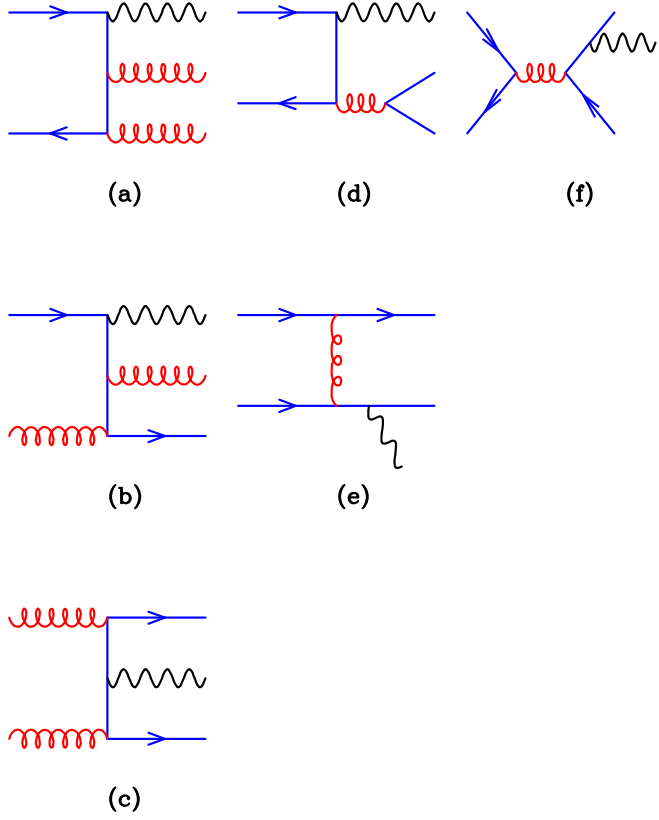}
\caption{\label{WZ2jet_born}Sample diagrams
for the process $\mbox{parton}+\mbox{parton} \to W/Z + 2~\mbox{partons}.$
As usual the vector boson is denoted by a wavy line.}
\end{figure}

\begin{figure}
\centering
\includegraphics[angle=270,width=8.6cm]{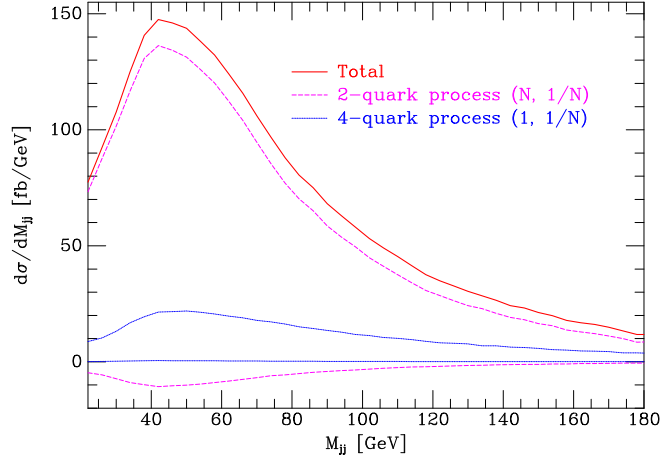}
\caption{\label{mjj}Color structure of the $W$+2 jet 
cross section vs. the dijet mass $M_{jj}$ at leading order.
Proceeding from the top the five curves are, 
the total LO result, 
the two quark process at $O(N)$, 
the four quark process at $O(1)$, 
the four quark process at $O(1/N)$ 
and the two quark process at $O(1/N)$,
where $N$ is the number of colors.}
\end{figure}

A representative sample of the Born diagrams for the processes,
\begin{equation}
\mbox{parton} + \mbox{parton} \rightarrow W/Z/\gamma^* + \mbox{2~partons}, 
\end{equation}
is shown in Fig.~\ref{WZ2jet_born}. In the coding of these processes
into MCFM we have made an artificial separation between processes involving
two quarks, Fig.~\ref{WZ2jet_born} (a-c)
and processes involving four quarks Fig.~\ref{WZ2jet_born} (d-f). 
This separation is motivated by the relative size of the contributions 
at leading order, illustrated in Fig.~\ref{mjj}, as well as by the relative
complexity of evaluating the different matrix elements.
The $2$-quark process at leading $N$ dominates the total, a trend that
we find is preserved at next-to-leading order.

We have not recalculated the virtual corrections to the 
basic Born process which are given by Bern, Dixon and 
Kosower in ref.~\cite{Bern:1997sc}.
These amplitudes have been calculated in the four dimensional helicity
scheme, which we consistently use throughout this program.

\begin{figure}
\centering
\includegraphics[width=8.6cm]{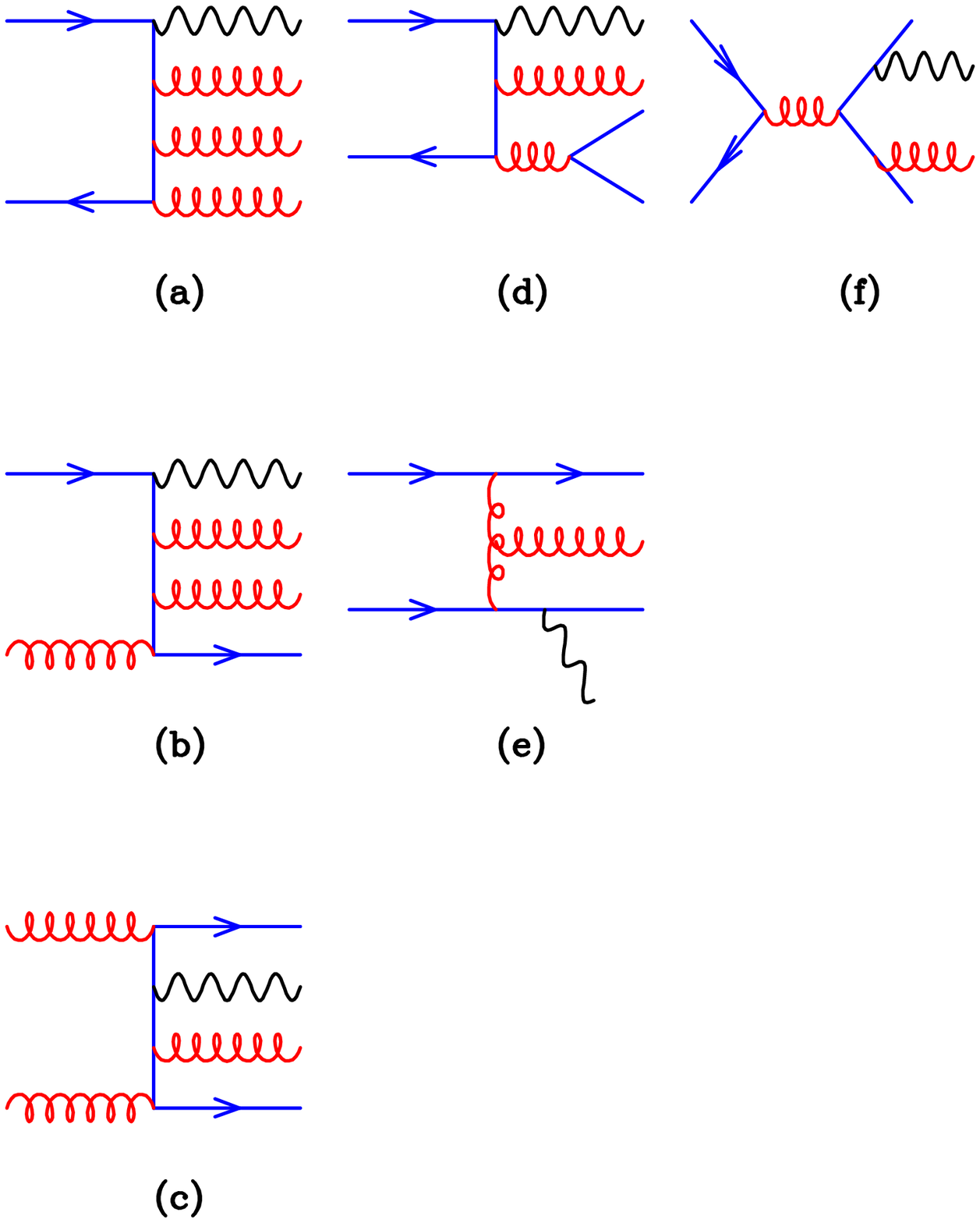}
\caption{\label{WZ2jet_real}Sample diagrams
for the process $\mbox{parton}+\mbox{parton} \to W/Z + 3~\mbox{partons}$.}
\end{figure}

The real corrections to the basic Born processes,
\ie~the processes,
\begin{equation}
\mbox{parton} + \mbox{parton} \rightarrow W/Z/\gamma^* + \mbox{3~partons} 
\end{equation}
have been published in \cite{Berends:1988yn,Hagiwara:1988pp,Nagy:1998bb}
and a representative sample of the contributing diagrams
is shown in Fig.~\ref{WZ2jet_real}.
These matrix elements are incorporated in MCFM using
the subtraction method of 
Ellis, Ross and Terrano~\cite{Ellis:nc,Ellis:1980wv}.
In this method, one constructs
counterterms having the same singularity structure as the real emission 
matrix elements, starting from the eikonal formula for soft emission.
The basic operating procedure is to create subtraction terms which
contain the same singularity structure as the lowest order diagrams,
but are simple enough that they are can be integrated over the phase 
space of the unobserved parton.
In order to improve the cancellation, the
kinematics of the counterterms are included using the prescription
of Catani and Seymour~\cite{Catani:1996vz}. For the case of 
$W +2$~jets there are 24 different counterterms, each with its 
own kinematic structure, (a {\it counterevent}). The kinematics of the
final state partons in the counterevents coincide with the 
kinematics of the event in the appropriate 
soft and collinear limits.

\subsection{Numerical checks}
The matrix elements for hadroproduction of $W/Z$+3 partons contain
many singularities which are subtracted 
by (dipole) counterterms. Although the enumeration of these counterterms
is in principle straightforward, the success of the whole program
depends on it being implemented correctly. We shall therefore present
a few details of the checks which we performed.

The real matrix elements were taken from ref.~\cite{Nagy:1998bb}
supplemented in certain cases by our own calculations. The coding of 
these matrix elements was checked by comparison with routines generated by the
program MADGRAPH~\cite{Stelzer:1994ta} powered by HELAS~\cite{Murayama:1992gi}.
We cannot use the routines generated by MADGRAPH directly 
because the resulting code is too slow to implement in 
a Monte Carlo program which requires many calls to the matrix element routine.

The next step is to verify that the numerical value of the
counterterms is in fact equal in magnitude to the real matrix element
in the singular limit. This is done by generating sets of points which lie
in all of the potentially singular regions and then checking the
cancellation of the event and the appropriate counterevent in each limit.

Finally, we must add back the counterterms, suitably integrated over the 
phase space of the emitted parton. Here it is clearly important
to add back exactly 
what has been subtracted in the previous step. We have tried to 
structure the code so that the comparison between the two steps 
is transparent. The integral of the counterterms 
over the emitted parton sub-space contains singularities,
which are regulated using dimensional regularization, in addition
to finite contributions. The code is structured
so that the finite parts are closely associated with the singular parts.
Thus the cancellation of the singular parts, 
(with the sum of the singular parts of the virtual
matrix elements and the Altarelli-Parisi factorization 
counterterms for the parton 
distributions), provides some assurance that the finite terms are included
correctly.

\section{Monte Carlo results}

\subsection{Input parameters}

MCFM has a number of default electroweak parameters which we use
throughout this paper. They are given in Table~\ref{ewparams}.
\begin{table}
\begin{center}
\begin{tabular}{|c|c|c|c|} \hline
Parameter & Default value & Parameter & Default value \cr 
\hline
$M_Z$        & 91.187 GeV  & $\alpha(M_Z)$     & 1/128.89 \cr
$\Gamma_Z$   & 2.49 GeV    & $G_F$             & 1.16639$\times$10$^{-5}$ \cr
$M_W$        & 80.41 GeV   & $g^2_w$           & 0.42662 (calculated) \cr
$\Gamma_W$   & 2.06 GeV    & $\sin^2 \theta_w$ & 0.23012 (calculated) \cr 
\hline
\end{tabular}
\label{default} 
\end{center}
\caption{Default parameters in the program MCFM.
\label{ewparams}}
\end{table}
As noted in the table, some parameters 
are calculated using the effective field theory
approach~\cite{Georgi:1991ci},
\begin{equation}
e^2 =  4 \pi \alpha(M_Z),\;\; g_w^2 =  8 M^2_W  \frac{G_F}{\sqrt{2}},\;\;
\sin \theta_w =  \frac{e^2}{g_w^2}.
\end{equation}
For simplicity we have taken the CKM matrix to be diagonal in 
the $W+2~{\rm jets}$ process. As a consequence there are, for example,
no $u{\bar s}$ initial states for this case. This approximation 
is not expected to influence any anticipated analyses.
For the other processes we retain only 
the Cabibbo sector of the CKM matrix,
\begin{equation}
V_{CKM}= \left( \begin{array}{ccc}  
     0.975 & 0.222 &  0 \\ 
     0.222 & 0.975 & 0  \\ 
     0 & 0 &        1    \\ 
\end{array} \right) \; .
\end{equation}

The value of $\alpha_S(M_Z)$ 
is not adjustable; it is determined by the
chosen parton distribution. 
A collection of modern parton distribution functions
is included with MCFM, but here we concentrate only on 
one of the MRST2001~\cite{Martin:2001es} sets with $\alpha_S(M_Z)=0.119$. 
We refer to this set as MRS0119, which is the label used
in our program.

\subsection{Basic cuts and jet selection}
\label{cutsdesc}

For all the results presented here, we consider only a positively charged
$W$ and choose the leptonic decays,
\begin{equation}
W^+ \to \nu e^+ , \qquad Z/\gamma^* \to e^- e^+. 
\end{equation}
In this paper we shall present results for the Tevatron collider only
and we pick a simple set of cuts accordingly. All leptons satisfy,
\begin{equation}
p_T^{\rm lepton} > 20~{\rm GeV}, \qquad |y^{\rm lepton}| < 1,
\end{equation}
and for the $W$ case there is also a cut on the missing transverse
momentum, $p_T^{\rm miss} > 20~{\rm GeV}$.
Our final requirement is that the dilepton mass
be greater than $15~{\rm GeV}$. Although this has no effect
in the $W$ case, it prevents the production of soft $e^- e^+$ pairs
which would otherwise be copiously produced by the virtual photon
in the $Z/\gamma^*$ process. 

Jets are found using the Run II $k_T$ clustering
algorithm~\cite{Blazey:2000qt} with
a pseudo-cone of size $R=0.7$, and are also subject to,
\begin{equation}
p_T^{\rm jet} > 15~{\rm GeV}, \qquad |y^{\rm jet}| < 2.
\end{equation}
For the new results on $W,Z + 2~{\rm jet}$ production, in this paper
we will mostly consider events where exactly 2 jets are
found by the algorithm, i.e. exclusive 2 jet production. The inclusive
production of jets - which would include events with 3 jets at
next-to-leading order - is a further option in MCFM that will only be
touched on briefly here.

\subsection{Scale dependence}
\label{scaledep}

The principle motivation for performing a next-to-leading order calculation
is to reduce the uncertainties in leading order predictions. In
particular, any perturbative prediction contains an unphysical
dependence on renormalization and factorization scales (often chosen to
be equal, as we shall do here). The magnitude of cross-sections and the
shape of differential distributions can vary greatly between two
different choices of scale, which is often interpreted as an inherent
``theoretical uncertainty'' which is then ascribed to the predictions.
Another strategy is to argue for a particular choice of scale, based
on the physics of the process under consideration.

\begin{figure*}
\centering
\includegraphics[angle=270,width=18cm]{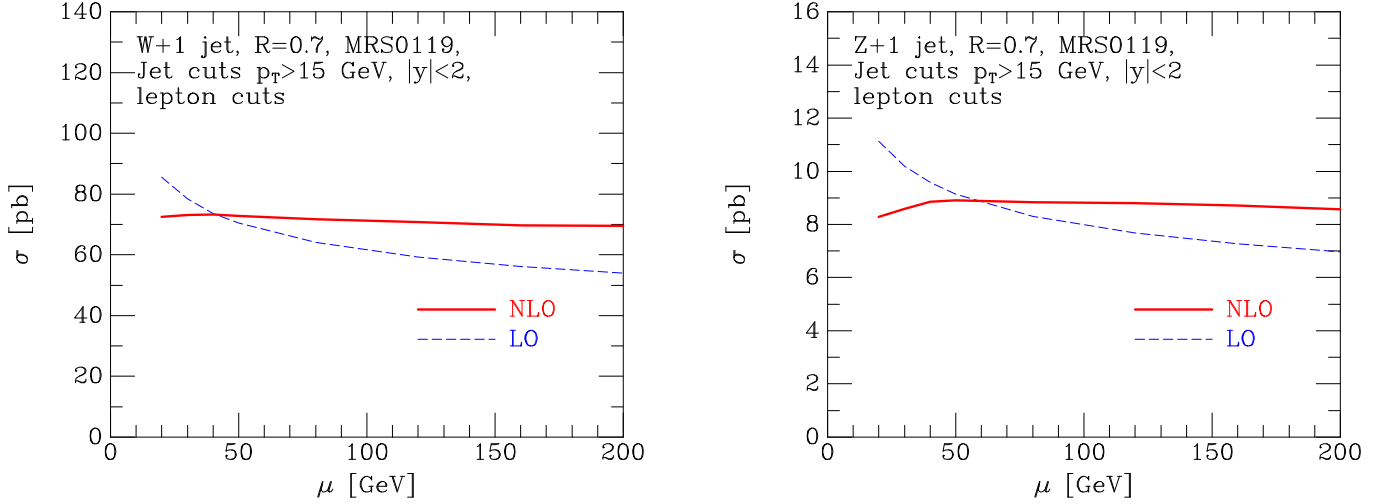}
\caption{\label{scaledep_wz1jet}
The scale dependence of the $W$ and $Z$+1~jet predictions,
with the factorization and renormalization scales equal and
given by $\mu$.
The differential distributions $d\sigma/dp_T$ 
are integrated over
$15 < p_T < 200~{\rm GeV}$, with the basic cuts as described in
section~\protect{\ref{cutsdesc}}.}
\end{figure*}
\begin{figure*}
\centering
\includegraphics[angle=270,width=18cm]{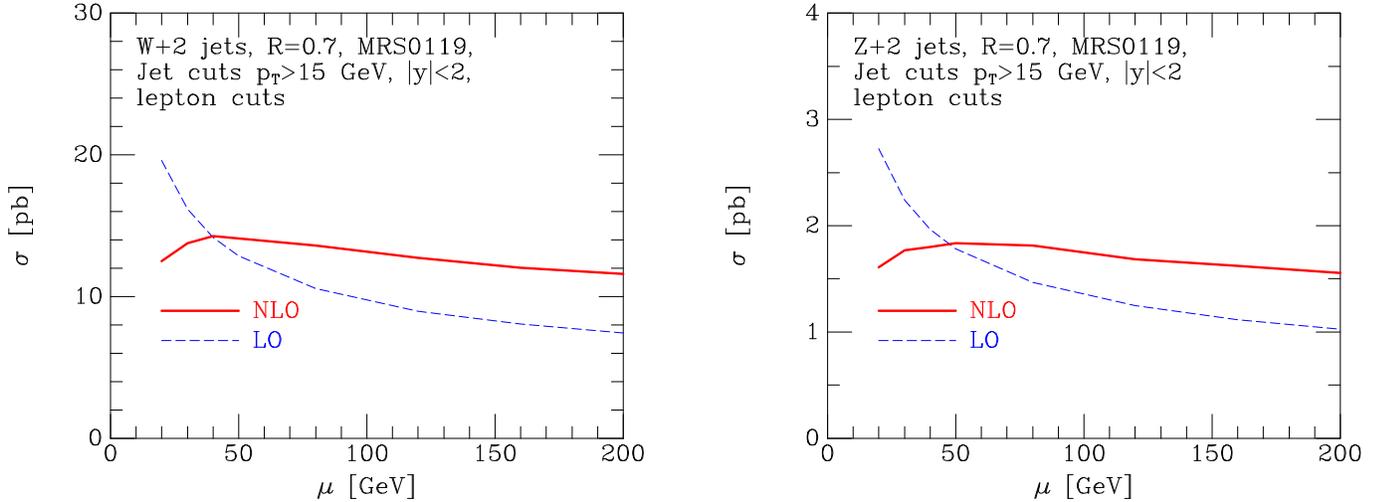}
\caption{The scale dependence of the $W$ and $Z$+2~jet
differential cross sections, $d\sigma/dM_{jj}$ integrated over
$20 < M_{jj}<200$ GeV. For more details, see section~\ref{scaledep}.}
\label{scaledep_wz2jet}
\end{figure*}
A next-to-leading order calculation is an invaluable tool for
investigating the issue of scale dependence. The logarithms that
are responsible for the large variations under changes of scale at
leading order are exactly canceled through to next-to-leading order. 
As a result, one expects that next-to-leading order predictions are 
more stable under such
variations. In addition, the next-to-leading order result may
provide further evidence to support a particular scale choice
that may have been deemed appropriate at leading order.

As an example of expected results, in Figure~\ref{scaledep_wz1jet}
we show the scale
dependence of the exclusive $W$ and $Z$ + 1 jet differential
cross-sections $d\sigma/dp_T$, integrated over the range
$15 < p_T < 200~{\rm GeV}$. The next-to-leading order predictions
have been known for some time~\cite{Giele:dj}, but here are
calculated within our program, MCFM.
For both processes, the leading order
prediction rises sharply as the scale is decreased, while the
corrections produce a far flatter curve that exhibits a much less
pronounced dependence on the scale choice.

The corresponding new results for the $2$-jet processes are shown in
Figure~\ref{scaledep_wz2jet}. As in the $1$-jet case, the
renormalization and factorization scales are set equal and the basic
cuts mentioned in the previous section are applied. In addition, we
now use the dijet mass differential distribution, $d\sigma/dM_{jj}$
integrated over $20 < M_{jj}<200$ GeV. As anticipated, both processes
show a considerable reduction in scale dependence. For example, the
ratio of the leading order prediction for the $W$ process using a hard
scale $\mu=2m_W$ to the result for a far softer scale $\mu=m_W/2$ is,
\begin{equation}
\frac{\sigma_{LO}(W+2~{\rm jets},\mu=m_W/2)}
     {\sigma_{LO}(W+2~{\rm jets},\mu=2m_W)}
 = 1.7,
\end{equation}
while the same ratio at next-to-leading order is only,
\begin{equation}
\frac{\sigma_{NLO}(W+2~{\rm jets},\mu=m_W/2)}
     {\sigma_{NLO}(W+2~{\rm jets},\mu=2m_W)}
 = 1.1.
\end{equation}

\subsection{$p_T$ distributions}

Once again, we repeat some $W,Z+1$~jet results, in order
to highlight both the similarities and the differences with the
corresponding $2$-jet distributions.

\begin{figure*}
\centering
\includegraphics[angle=270,width=18cm]{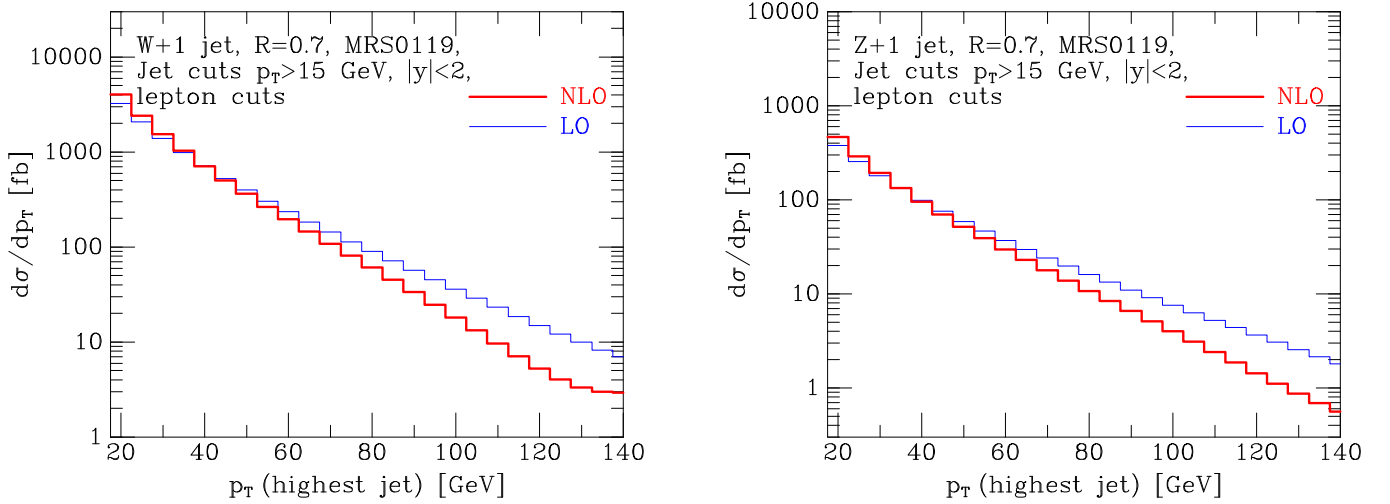}
\caption{The jet $p_T$ distribution of $W$ and $Z$+1~jet events,
evaluated with the hard scale choice, $\mu=80$~GeV.}
\label{ptdep_wz1jet}
\end{figure*}
In Figure~\ref{ptdep_wz1jet} we show the jet $p_T$ distribution
for both of the $1$-jet cases, using a relatively hard choice of
scale, $\mu=80$~GeV. We first note that since we are
considering the exclusive jet cross-section, the rise of the
distributions at low $p_T$ is limited only by the jet cut,
$p_T > 15$~GeV. At next-to-leading order the distributions
change significantly to become much softer. 
At high $p_T$ a single
jet is much more likely to radiate a soft parton (that passes the
fixed $p_T$ cut and is counted as an extra jet), thus removing it from
the sample~\cite{Giele:dj}.

\begin{figure*}
\centering
\includegraphics[angle=270,width=18cm]{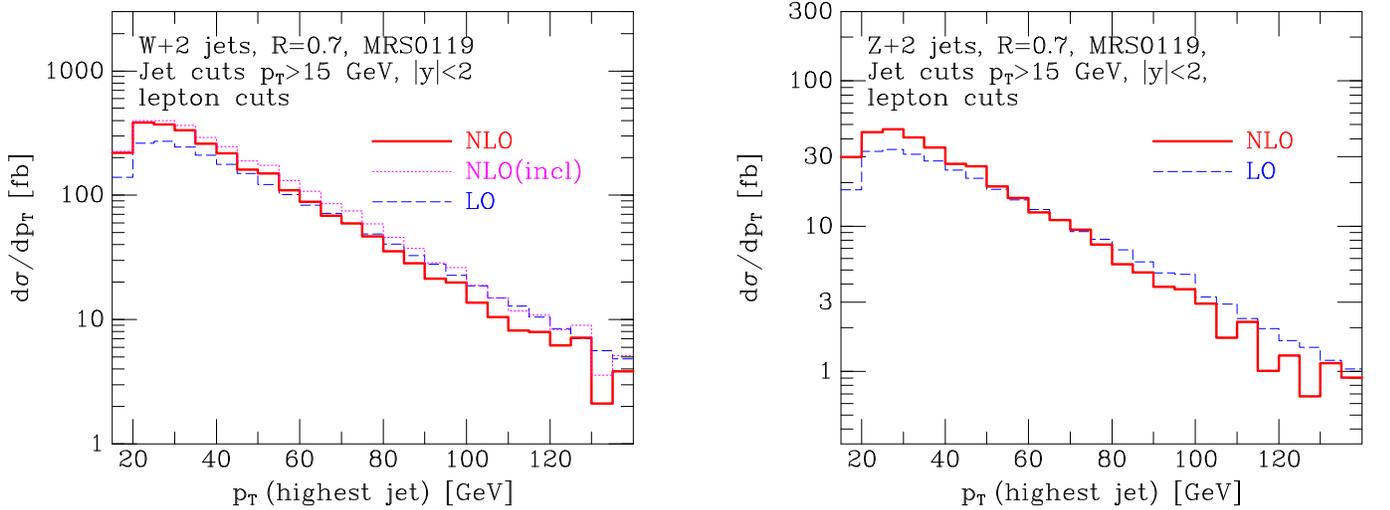}
%\special{psfile=z2jet_pt5.ps 
%angle=270 hscale=50 vscale=50 hoffset=-10 voffset=270}
\caption{The hardest jet $p_T$ distribution in
$W$ and $Z$+2~jet events, using the scale $\mu=80$~GeV.}
\label{ptdep_wz2jet}
\end{figure*}
The situation for the $2$-jet processes is shown in
Figure~\ref{ptdep_wz2jet}, where we plot the $p_T$ distribution of
the hardest jet, again using $\mu=80$~GeV. In contrast with the
previous figure, the distributions turn over at small $p_T$.
If the highest $p_T$ jet has a $p_T \leq 20$~GeV,
there is little phase space for the emission of a second softer jet
with $p_T > p_T^{min} =15$~GeV.
We also see that including the radiative corrections softens
the distribution considerably, for the same reason as before.
For the $W$ case, we also show the {\it inclusive}
distribution, \ie~the cross section for the production of two or more jets.
This `fills in' the high-$E_T$
tail of the distribution.

\section{Heavy flavor content of jets}

\subsection{$\mbox{Hadron} + \mbox{Hadron} \rightarrow W/Z + b + \bar{b} $}

\begin{figure}
\centering
\includegraphics[width=8.6cm]{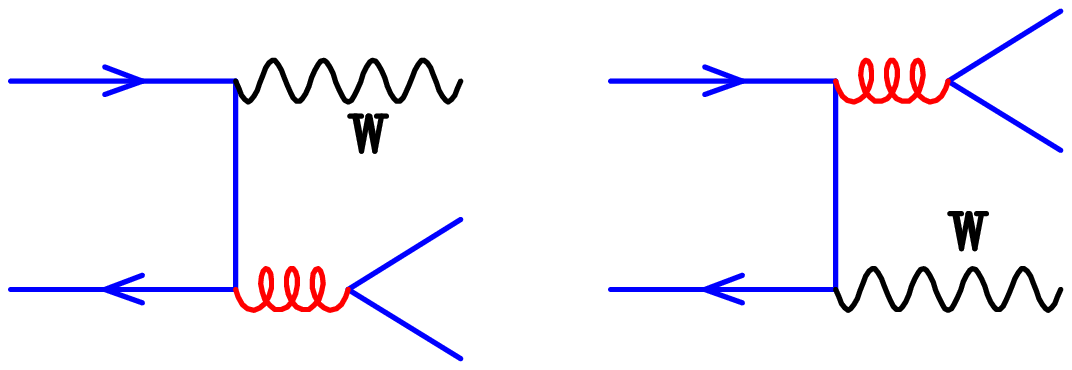}
\caption{\label{Wbb_born}Lowest order diagrams
for the process $\mbox{parton}+\mbox{parton}\to W b \bar{b}$.}
\end{figure}

We would like to estimate the fraction of $W+2$~jet events that
contain two heavy quark jets.  We will limit our discussion to $b$
quarks, because they can be tagged with high efficiency. In order to
do so, we recall the next-to-leading order results for the production
of a $b \bar{b}$ in association with a $W$, reported
in~\cite{Ellis:1998fv}. As a reminder to the reader, we work in the
approximation in which the $b$ quarks are taken to be massless and we
have ignored contributions from processes in which there are two $b$
quarks already present in the initial state. The basic lowest order
diagrams for this process are shown in Fig.~\ref{Wbb_born}.
$Wb\bar{b}$ processes accompanied by up to 4 jets have been considered
at tree level in ref.~\cite{Mangano:2001xp}.

\begin{figure}
\centering
\includegraphics[width=8.6cm]{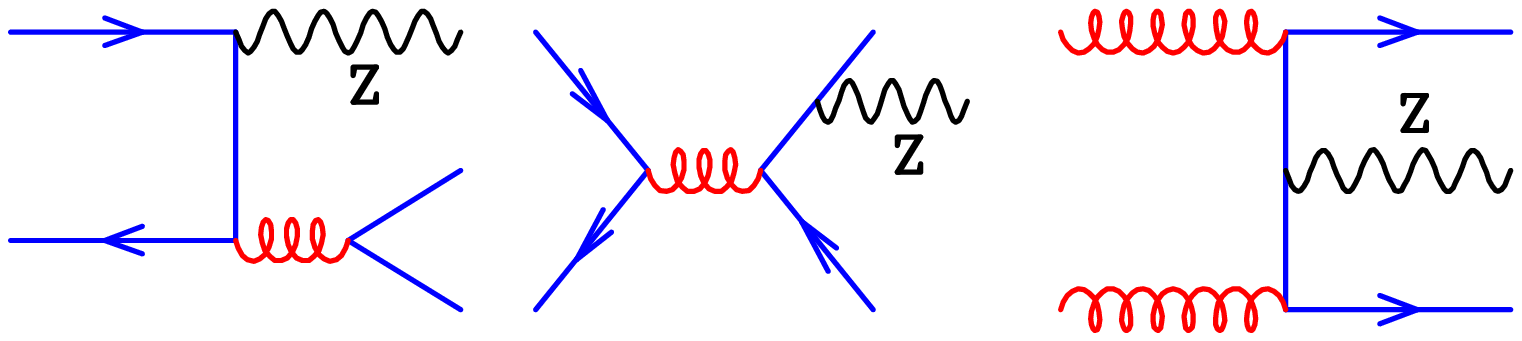}
\caption{\label{Zbb_born}Sample diagrams for the process 
$\mbox{parton}+\mbox{parton}\to Z b \bar{b}$.}
\end{figure}

For the related study including a $Z$ instead of a $W$, we use the
results presented in~\cite{Campbell:2000bg} for the production of a
$b \bar{b}$ pair in association with a $Z$. The same approximations
apply as discussed above for the $W$ case. The notable difference
now is that there are more lowest order diagrams, as shown in
Fig.~\ref{Zbb_born}, including an initial state composed only of gluons.
As discussed in~\cite{Campbell:2000bg}, these latter diagrams with initial 
gluons are believed to be responsible for the sizeable corrections to the
basic process at large $m_{b\bar{b}}$.

\begin{figure*}
\centering
\includegraphics[angle=270,width=18cm]{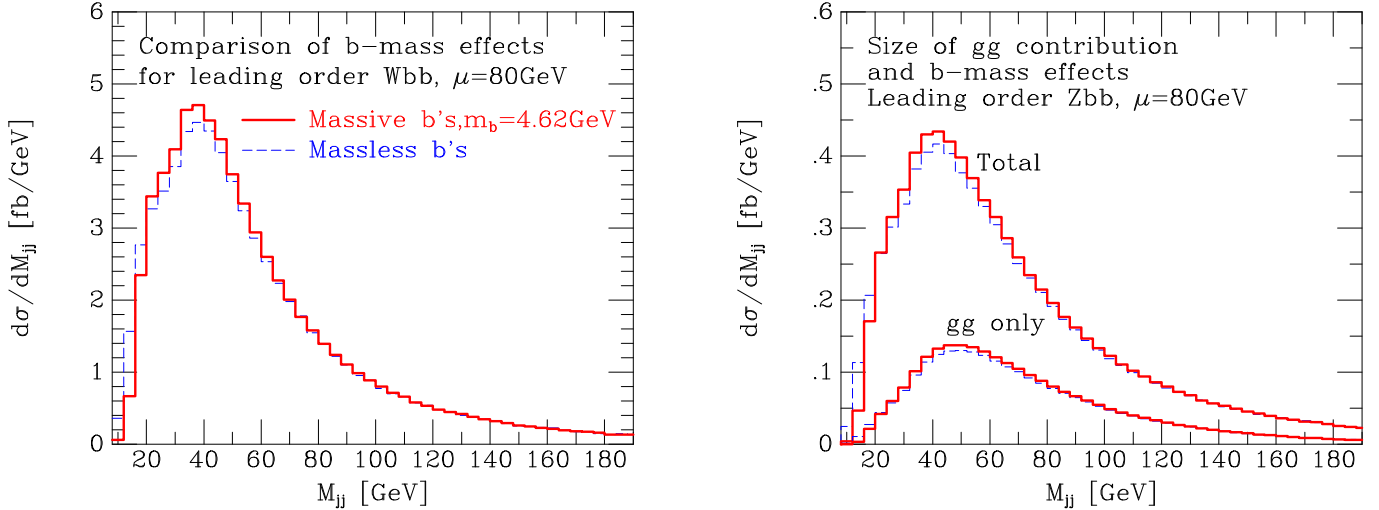}
\caption{\label{comp_wbbbar_mass}Illustration of mass effects in lowest order.
$M_{jj}$ is the mass of two tagged $b$-jets.}
\end{figure*}

An immediate concern is that neglecting the $b$-quark mass may be
unjustified~\cite{Mangano:1992kp}. 
At low values of $m_{b \bar{b}}$, quark mass effects
may be important.  To address these concerns, in
Figure~\ref{comp_wbbbar_mass} we compare the lowest order $m_{b \bar{b}}$ 
distribution calculated using the full mass dependence with the
result obtained by setting $m_b=0$. There are two effects of
introducing a mass for the $b$-quark. Firstly, the phase-space becomes
smaller, leading to a reduction of the cross-section.  On the other
hand, the matrix elements receive extra contributions proportional to
powers of $m_b^2$ which may increase the result.  As shown in
Figure~\ref{comp_wbbbar_mass}, the matrix element effects dominate around
the peak of the distribution where they are as large as 5\%. Closer to
threshold the phase space effects are dominant.  At large $m_{b \bar{b}}$ 
the quark mass effects are quite small as expected.

\subsection{Results}

\begin{figure*}
\centering
\includegraphics[angle=270,width=18cm]{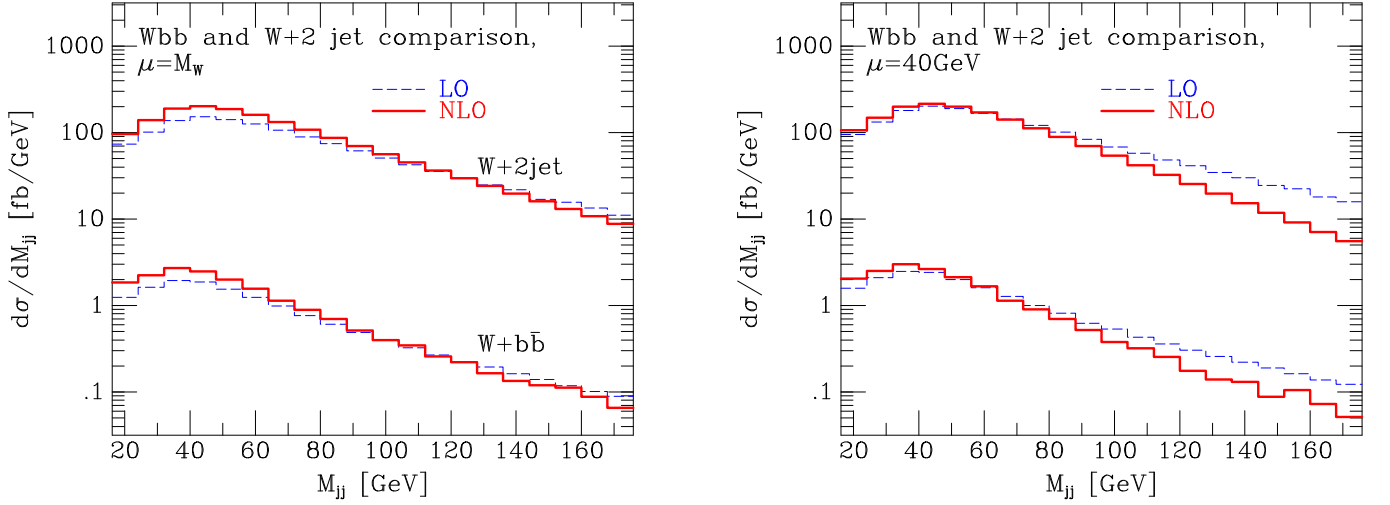}
%\special{psfile=comp_wbbbar_w_2jet_40.ps
%angle=270 hscale=50 vscale=50 hoffset=-10 voffset=270}
\caption{\label{comp_wbbbar_w_2jet}Comparison of the expected 
distributions in the
dijet mass for untagged $W$+2~jet events and
$W$+2~$b$-tagged jet events. A hard scale, $\mu=80~{\rm GeV}$,
is shown in the left-hand plot and the softer scale,
$\mu=40~{\rm GeV}$, on the right.}
\end{figure*}

\begin{figure*}
\centering
\includegraphics[angle=270,width=18cm]{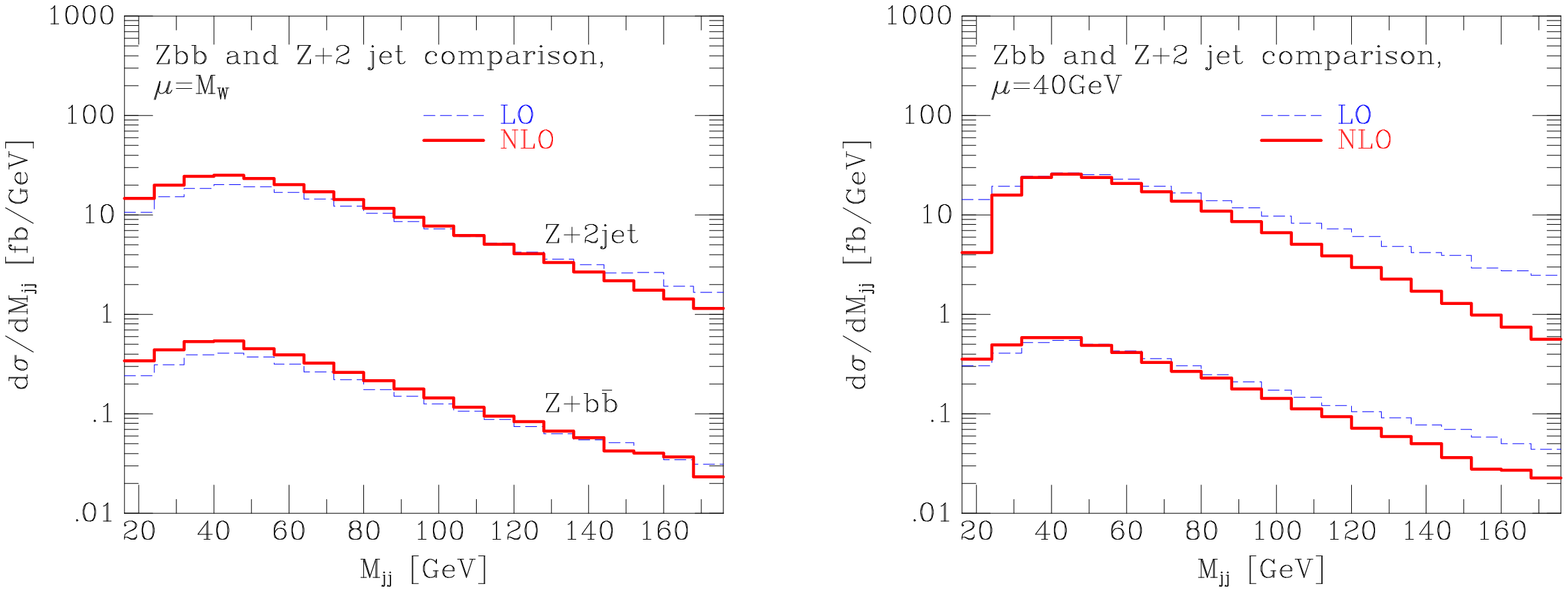}
\caption{\label{comp_zbbbar_z_2jet}Comparison of the expected 
distributions in the
dijet mass for untagged $Z$+2~jet events and
$Z$+2~$b$-tagged jet events. A hard scale, $\mu=80~{\rm GeV}$,
is shown in the left-hand plot and the softer scale,
$\mu=40~{\rm GeV}$, on the right.} 
\end{figure*}

In order to compare the results for $b$-quark jets with those for the
whole $2$-jet sample, we will show the differential
cross-section as a function of the dijet mass.
We use two choices of scale in
these analyses, a hard scale $\mu \approx M_W, M_Z = 80$~GeV
and a softer scale $\mu = 40$~GeV. These are the
scales used for the plots shown in
Figures~\ref{comp_wbbbar_w_2jet}~and~\ref{comp_zbbbar_z_2jet},
where both the leading order and the radiative corrections are shown
for comparison.
\begin{figure*}
\centering
\includegraphics[angle=270,width=18cm]{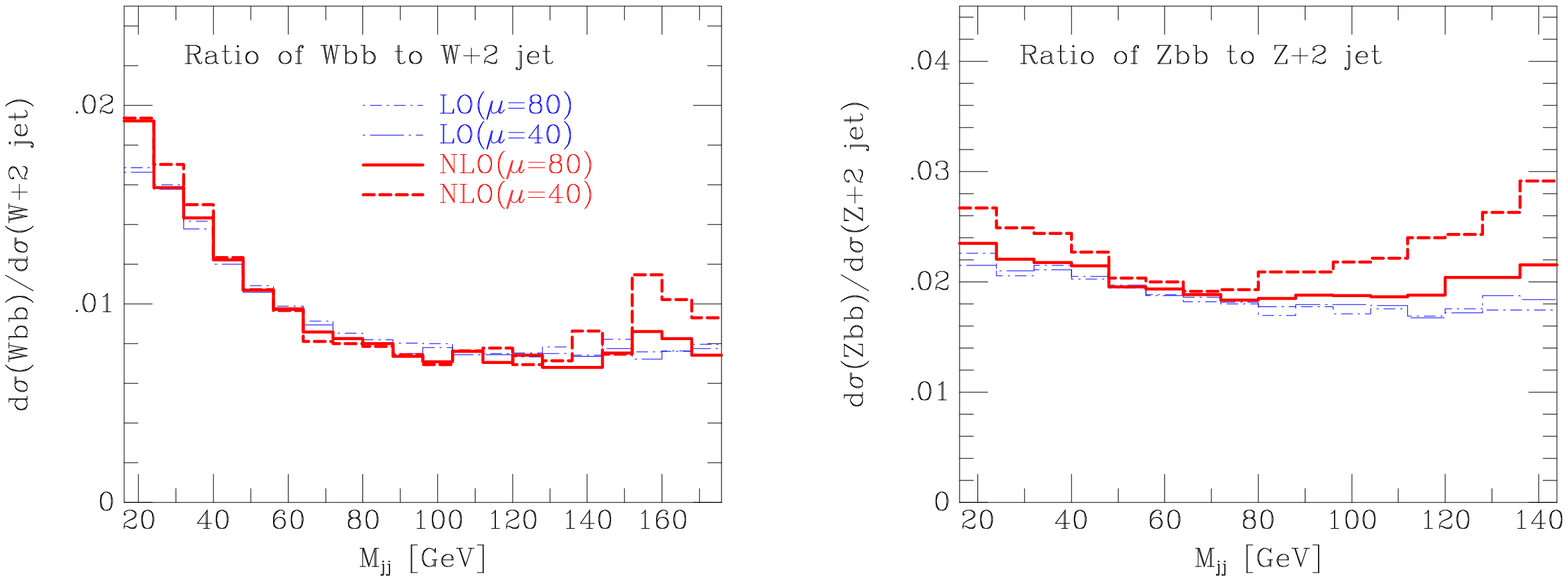}
%\special{psfile=bb_ratioz.ps
%angle=270 hscale=50 vscale=50 hoffset=-10 voffset=270}
\caption{\label{bb_ratio}Ratio of $W/Z$+2~b-tagged jets
to $W/Z+2$~jet events in LO and NLO at both $\mu=40$~GeV
and $\mu=80$~GeV.}
\end{figure*}

As with the $p_T$ distributions of the previous section, the
shapes of the distributions change when the QCD corrections are included.
For both the $b{\bar b}$ and general $2$~jet distributions, the hard scale
causes the dijet cross-section to increase at next-to-leading order
for small values of $M_{jj}$. The radiative
corrections using the soft scale cause a considerable depletion
in the cross-section at high $M_{jj}$. However, the shapes of the
$b{\bar b}$ and $2$~jet distributions appear very similar when
compared at the same order of perturbation theory and using the same scale.

In Fig.~\ref{bb_ratio} we show the cross section for events that contain
$2$ $b$-tags divided by the cross section for all two jet events, 
as a function of the dijet mass. As can be seen,
for the $W$ this ratio does not depend very strongly on either the
choice of scale or the order in perturbation theory. The
percentage falls at low values of $M_{jj}$
until $M_{jj} \sim 60$~GeV, where it becomes approximately constant
at $0.8\%$.
For the $Z$, the proportion is fairly constant at approximately
$2\%$ for all the curves except for the case of next-to-leading order
at $\mu=40$~GeV. In this case the percentage rises at high $M_{jj}$.
The origin of this effect may be associated with the extra diagrams
present in the $Z$ case and requires further study.

\section{Conclusions}

We have presented the first results for the implementation of
$W/Z+2~$ jet production at next-to-leading order in a general
purpose Monte Carlo. An analysis based on exclusive jet
production for Run II of the Tevatron
shows that the usual benefits of next-to-leading order are
realized, among them being a reduced scale dependence and hence an
improved normalization for distributions. We also find changes in
the shapes of distributions similar to those found in the $1$-jet
case. These modifications are reduced if we 
consider the inclusive cross-section.

We performed an  analysis of the heavy flavor content of
jets produced in association with a vector boson. For production in
association with a $W$, the ratio of $b$-tagged to untagged jets
changes very little upon the inclusion of radiative corrections and
appears to be predicted very well by perturbation theory. 

\begin{acknowledgments}
We would like to thank the Fermilab Computing Division for computer
time on the fixed target farm.
This work was supported in part by the U.S. Department of Energy
under Contracts No. W-31-109-ENG-38 (Argonne) and
No. DE-AC02-76CH03000 (Fermilab).
\end{acknowledgments}

\clearpage

\end{document}